\documentclass[journal,twoside,web]{ieeecolor}
\usepackage{generic}
\usepackage{cite}
\usepackage{amsmath,amssymb,amsfonts}
\usepackage{algorithmic}
\usepackage{bm}
\usepackage{graphicx}
\usepackage{textcomp}
\usepackage{url}
\def\BibTeX{{\rm B\kern-.05em{\sc i\kern-.025em b}\kern-.08em
    T\kern-.1667em\lower.7ex\hbox{E}\kern-.125emX}}
\begin{document}
\title{Detection and Localization of Load Redistribution Attacks on Large Scale Systems}
\author{Andrea Pinceti, \IEEEmembership{Student, IEEE}, Lalitha Sankar, \IEEEmembership{Senior Member, IEEE}, and Oliver Kosut, \IEEEmembership{Member, IEEE}
\thanks{The authors are with the School of Electrical, Computer and Energy Engineering at Arizona State University, Tempe, AZ, USA. Their emails are: apinceti@asu.edu, lalitha.sankar@asu.edu, and oliver.kosut@asu.edu.}
\thanks{We would like to thank Mr. Zhigang Chu at ASU for providing access to the attack design code. This material is based upon work supported by the National Science Foundation under Grants No. CNS-1449080 and OAC-1934766 and the Power System Engineering Research Center (PSERC) under projects S-72 and S-87.}
}

\maketitle

\begin{abstract}
A nearest neighbor-based detection scheme against load redistribution attacks is presented. The detector is designed to scale from small to very large systems while guaranteeing consistent detection performance. Extensive testing is performed on a realistic, large scale system to evaluate the performance of the proposed detector against a wide range of attacks, from simple random noise attacks to sophisticated load redistribution attacks. The detection capability is analyzed against different attack parameters to evaluate its sensitivity. Finally, a statistical test that leverages the proposed detection algorithm is introduced to identify which loads are likely to have been maliciously modified, thus, localizing the attack subgraph. This test is based on ascribing to each load a risk measure (probability of being attacked) and then computing the best posterior likelihood that minimizes log-loss.
\end{abstract}


\section{Introduction}
\label{sec:introduction}

\IEEEPARstart{T}{he} electrical grid is a constantly evolving cyber-physical system and, as such, it is increasingly reliant on information and communication technology. A vast research effort undertaken in the past decade in the field of cybersecurity of power systems has identified some crucial vulnerabilities of the cyber layer which can be exploited to disrupt the physical system. In this context, \cite{Liu2009} shows that state estimation (SE) and the traditional bad data detector (BDD) used in energy management systems (EMSs) can be easily spoofed and bypassed via false data injection (FDI) attacks. This finding represents the basis for the design of a wide class of attacks called load redistribution (LR) attacks. Load redistribution attacks can be performed by injecting intelligently designed false measurements that lead to a wrong estimate of the system state. From the operator's perspective the attack makes it appear as if the system loads have changed from their actual values, without changing the net load.


In \cite{Yuan2011} and \cite{Zhang}, the concept of load redistribution attacks is formalized by developing a bi-level attacker-defender problem for targeted attacks. In this setting, the attacker can design false measurements which can cause physical consequences on the system; specifically, \cite{Zhang} attempts to find an attack, unobservable to the EMS, to cause a power overload on a target line. In a similar fashion, \cite{Sanjab,Xie} present examples of LR attacks on the electricity market: the authors show that it is possible to launch load redistribution attacks that create system congestion, thus manipulating locational marginal prices. 

We propose a new bad data detector that can identify LR attacks based on the analysis of load estimates, thus overcoming the limitations of SE and the traditional BDD. In \cite{Pinceti}, we developed three anomaly detectors, each based on a different machine learning technique: replicator neural network, support vector machine, and nearest neighbor. These detectors can effectively determine if the observed loads represent a normative system state or if they have been maliciously modified. The nearest neighbor-based detector works by finding in the historical data the closest load vector to the real time loads and, based on the measured Euclidean distance, a thresholding technique is used to decide if the loads are normative or anomalous. From our tests, nearest neighbor demonstrated the best performance out of the three proposed. 
In this paper, we build on this preliminary work to design an improved detector and an attack localization scheme. The novel contributions of this paper are as follows:
\begin{itemize}
	\item The basic detector is modified so that \textit{it scales to much larger power system models} while preserving the good detection performance shown in \cite{Pinceti}. This is achieved by devising a grouping strategy to organize the system loads into clusters that can be analyzed independently.
	\item \textit{Extensive testing and sensitivity analysis is performed to evaluate the performance of the detector} against intelligently designed LR attacks as well as random anomalous load changes. This allows for the characterization of the strengths and limitations of the detector. Furthermore, \textit{the proposed algorithm is integrated within a complete EMS platform} to showcase its detection performance and computational efficiency.
	\item Building on the improved detector, \textit{a statistical approach is presented to localize the attacks and determine the likelihood of each load of being attacked}. The deviation in loads is captured via a Z-score and log-loss is used as a measure to find the likelihood function that minimizes the error. This represents a crucial step towards the development of decision tools that can help operators to securely manage power systems when targeted by cyber-attacks.
\end{itemize}

Related work on the design of FDI and LR attack detectors can be found in the literature. For example, in   \cite{Joshi2017}, multiple linear regression is used to study the voltage profiles in a system and determine if a LR attack is taking place. Unlike our work, the method there proposed is designed for distribution systems and, as we will explain later, the attacks tested are not realistic as they simulate changes in loads of up to 100\%. Other work focuses on using deep neural networks to learn the temporal correlation which exists between the real-time measurements and previous samples \cite{Yu2018}, or verifying the statistical behavior of the estimated states over time \cite{Huang2016}. The assumption on which these detectors are built is that when an attack is injected, the false measurements are not compatible with the dynamics observed from the previous measurements thus making it possible to flag them as attacked. Based on this, a slow ramping attack which only slightly changes the system's state at each sampling time will likely not be detected.
Moreover, these detectors are tested on limited attack scenarios and their performance is not verified against multiple classes of attacks. Finally, while many attack detectors have been proposed, to the best of the authors' knowledge, the idea of detecting FDI attacks by identifying patterns in the observed loads has not been explored before.

%

The rest of the paper is organized as follows: in Section \ref{attmodel} a description of load redistribution attacks and how to design them is presented. In Section \ref{base} we summarize the basic detection algorithm we have presented in \cite{Pinceti} and show its performance limitations when used on large scale systems. The required improvements are described in Section \ref{large} and the detection results on a wide range of load redistribution attacks are presented in Section \ref{test}. Section \ref{localization} illustrates the statistical analysis that leverages the improved nearest neighbor-based detector to determine the buses that have been attacked.

\section{Attack model and design}\label{attmodel}
For a power system, the relationship between the vector of measurements $\bm{z}$ and the state vector $\bm{x}$ can be written as 
\begin{equation}
\bm{z}=h(\bm{x})+\bm{e}\label{se}
\end{equation}
where $h$ is the non-linear relationship between measurements and states (usually, complex bus voltages), while vector $\bm{e}$ represents the random measurement noise. As shown in \cite{Liu2009}, an unobservable attack can be constructed by replacing the original measurements $\bm{z}$ with a corrupted set of measurements $\bar{\bm{z}}$ such that
\begin{equation}\bar{\bm{z}}=h(x+c)\label{attack}\end{equation}
where $\bm{c}$ is the state attack vector. 
Based on this fundamental result, the authors in \cite{Zhang} present a bi-level optimization problem to compute an attack vector $\bm{c}$ that will maximize the power flow on a specific target line. To cause such physical consequences on the system, the false measurements must be designed in such a way that they will initiate a system response in the form of generation redispatch. This can be done by creating an unobservable attack that will lead state estimation to wrongly estimate the system loads, thus causing a wrong dispatch solution. The bi-level optimization problem proposed in \cite{Zhang} is improved in \cite{Chu} to make it more efficient and scalable to large scale systems. The first level models the attacker's choice of attack to maximize the overload on a target line; the second level models the system's response to the attack via a DCOPF to observe the resulting physical consequences. In designing the false measurements, the attacker is limited on how much the false loads can deviate from the real loads: the load shift factor represents the maximum percentage by which any load can be modified. This constraint comes from the fact that an operator would easily identify a large change in load over a short period of time as an anomaly; in the existing literature, a load shift of 20\% is considered the maximum allowable for an attack to remain unobservable. The attack detector presented in this paper aims at identifying in real time if the set of measured loads is genuine or if it is the result of an attack on state estimation. As we will show, the proposed detector is able to easily identify attacks with load shifts of 15\% and lower.

\section{Basic detection algorithm}\label{base}
\subsection{Small systems}
The proposed attack detection mechanism works by analyzing the correlation structure within the currently observed load values and comparing it to attack-free historical load data. The measured load configuration to be tested is given as input to the detector which generates a scalar value. This value is then compared against a threshold $\tau$ to label the loads as \textit{normative} or \textit{attacked}. To evaluate the detection performance, two metrics are used: \textit{detection probability}, which is the ratio between the number of cases correctly labelled as attacked and the total number of attacked cases tested, and \textit{false alarm rate}, which is the number of normative cases that are labelled as attacked divided by the total number of normative cases tested. The specific value of the threshold is chosen as a tradeoff between detection probability and false alarm rate. Our approach can be considered a semi-supervised learning problem since the detectors are trained only on normative data which is already widely available to operators. Because no attacked data is needed in the training phase, the detectors will not be biased towards specific types of attacks. 
Given the almost identical detection capability of the three detectors tested in \cite{Pinceti}, in the following work, the nearest neighbor detector is chosen for its computational and explanatory simplicity.

Nearest neighbor algorithms are based on the assumption that data labelled as normative lies in limited, dense regions of space while anomalies are located further from these neighborhoods \cite{nn1,nn2}. Let us define $\bm{p} \in \mathbb{R}^n$ as the vector of observed load values to be tested, where $n$ is the number of loads in the system. The normative data is represented by the set $\bm{P}_N^\text{hist} \in \mathbb{R}^{n \times n_h}$ of historical load vectors $\bm{h}_i \in \mathbb{R}^n$ that have been observed in the past, where $n_h$ is the total number of historical vectors.
The classification is done by measuring the Euclidean distance between the current load profile $\bm{p}$ and every vector $\bm{h}_i$ in the historical dataset (assumed to be attack-free). The closest distance $d$ for sample $\bm{p}$ is defined as
\begin{equation}
d=\min_{j=[1:n\textsubscript{\textit{h}}]} \| \bm{p}-\bm{h}_j\|\textsubscript{\textsubscript{2}}\label{distance}.
\end{equation}
To label $\bm{p}$ as normative or attacked, $d$ is compared against a predetermined threshold $\tau$.

\begin{figure}
	\centerline{\includegraphics[scale=0.4]{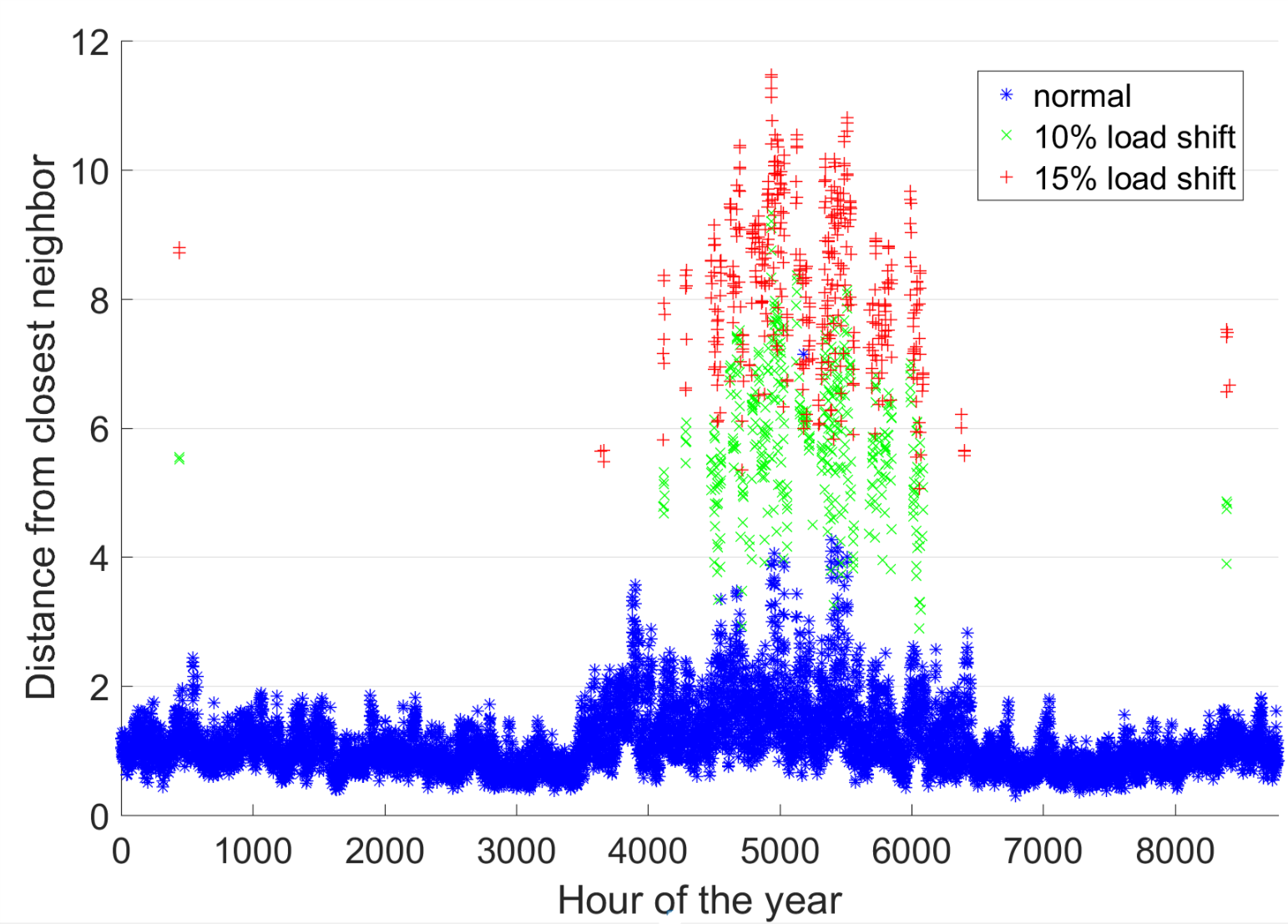}}
	\caption{IEEE 30 bus system: distribution of nearest neighbor distance for normative and attacked cases.}
	\label{NNscatter} 
\end{figure}

In \cite{Pinceti}, we tested this approach on the IEEE 30 bus system. Publicly available zonal historical load data from the PJM system \cite{PJM} was mapped to the loads of the 30 bus system to create hourly load profiles for 5 consecutive years. The proposed detector showed very high detection capability with low false alarm rates. Figure~\ref{NNscatter} is taken from \cite{Pinceti} and it shows some of the results obtained on this small system. The blue points represent the minimum distance for the normative load vectors (not attacked) while in green and red are the distances corresponding to attacked cases with 10\% and 15\% load shift, respectively. This illustrates how loads resulting from attacks lead to much higher nearest neighbor distances compared to normative load profiles; thus, suggesting that the minimum distance is an effective metric for attack detection. 

\subsection{Large systems}\label{largesystems}
\begin{figure}
	\centerline{\includegraphics[scale=0.5]{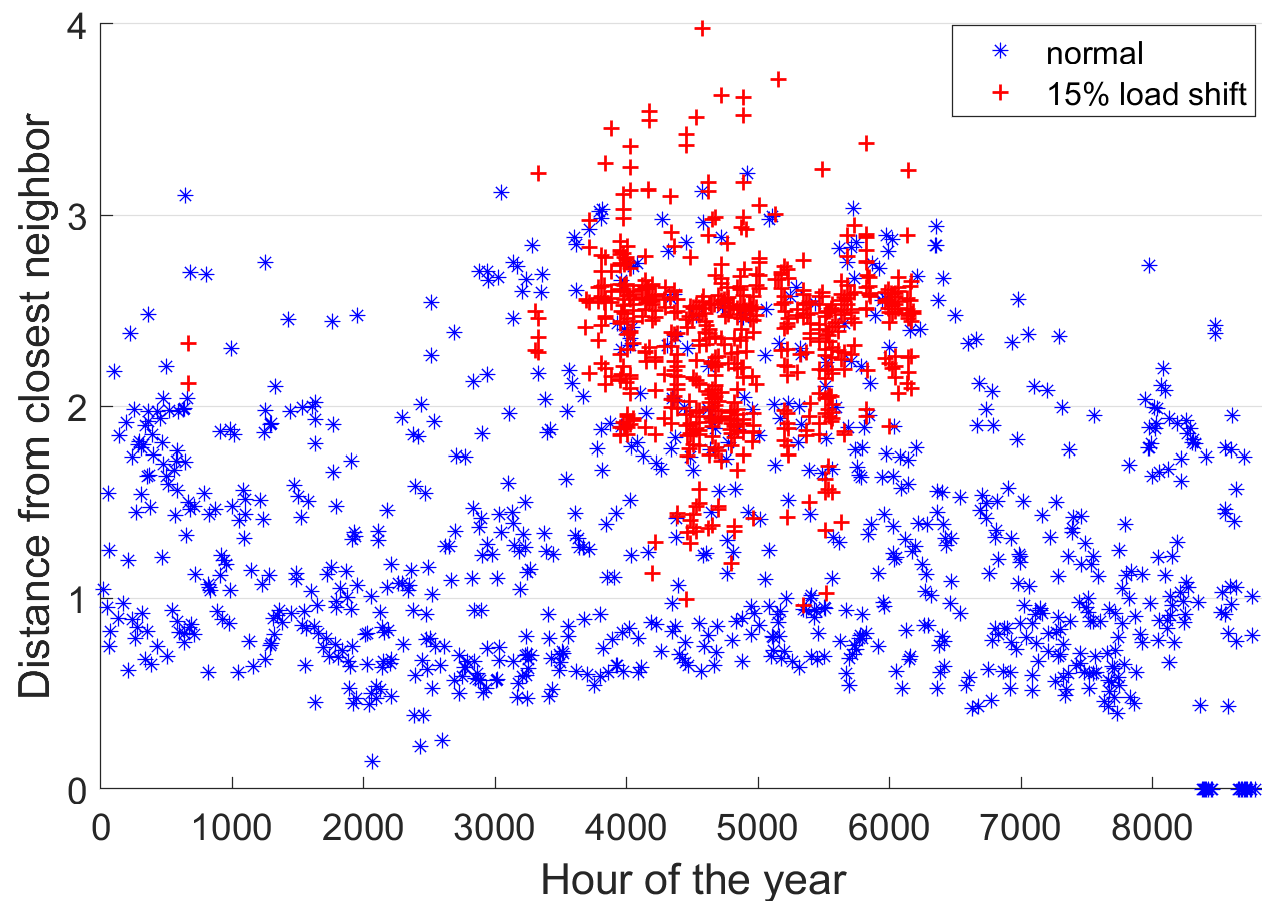}}
	\caption{Synthetic Texas system: distribution of nearest neighbor distance for normative and attacked cases.}
	\label{noGroups} 
\end{figure}

While the results obtained on the 30 bus system are promising, the detector needs to be tested on large scale systems to verify its performance in a more realistic setting and to guarantee its suitability for implementation in real system operations. To this end, the same analysis presented in \cite{Pinceti} and summarized in the previous section is performed on the synthetic Texas system \cite{Birchfield,Li2018}. This model, developed at Texas A\&M, is a synthetic grid of the state of Texas. It has 2000 buses, 3206 branches, and 1125 loads and it includes bus-level hourly load data for the year 2016. Using the attack model described in Section \ref{attmodel}, around 280 attacks with load shift of 15\% have been designed on the most congested cases. We randomly selected 90\% of the 8784 normative load vectors to represent the historical data, and the remaining 10\% for testing. The nearest neighbor algorithm is used to compute the minimum distance for the test and the attacked load vectors against the historical load data. Figure~\ref{noGroups} shows the minimum distance for each normative load vector (blue points) and for the attacked cases (red points). It is easy to see that the detector does not perform well, and that the attacked cases are indistinguishable from the normative ones. This can be explained by the fact that when measuring the Euclidean distance between two high-dimensional vectors, the contribution of a limited subset of dimensions is small. That is, if only a few tens of loads are attacked, the total distance measured over hundreds of loads will deviate only slightly from the distance computed on the load vector where no loads are modified. In this case, each load vector has dimension 1125 and the attacks modify only about 100 to 200 loads; the effect of the attacked loads is not large enough to result in distance values significantly higher than those of the normative data. 

\section{Detection on large scale systems}\label{large}
The simple test presented in the previous section shows that the basic nearest neighbor detector introduced in \cite{Pinceti} does not perform as well when applied to large scale systems (hundreds or thousands of buses). For this reason we need a new approach to improve the detection mechanism to be effective for any system, regardless of its size. The method we propose aims to leverage the capability of the nearest neighbor algorithm to identify anomalous loads even when only a small fraction of the total system loads are being attacked. 

Previous work has shown that in a large transmission-level system, load redistribution attacks tend to target only some portions of the network. As a consequence, the loads which are modified represent a subset of the total system loads and they are restricted to a subgraph of limited size. Based on these observations, the detection algorithm is modified so that it analyzes multiple predefined subsets of the system loads. In this work we propose a grouping strategy that can be used to divide the loads into relatively small groups such that they can be analyzed independently and in parallel by the attack detector. It is important to notice that the strategy presented in the next section is just one example of grouping which empirically worked well for the systems tested. Different grouping techniques, perhaps leveraging specific knowledge and insights regarding the power system to be secured, can be easily implemented within the framework of the proposed detection algorithm. 

\subsection{Grouping strategy}\label{grouping}
The first step required to define the load groups is to sort the loads based on their MW rating, from largest to smallest. Starting from the largest load, the first group is created by including the load itself and all its neighboring loads within a certain radius $r_g$, where the radius is measured as the smallest number of branches connecting two loads. At this point, the next largest load is selected and if it is not contained in any of the previous groups, a new group is created. This process is repeated until $n_g$ groups are defined. Note that it is possible for a bus to be contained in multiple groups, or no groups. The parameters $r_g$ and $n_g$ have a direct effect on the detection performance and their selection will be discussed in the next sections. As our results show, this grouping strategy proves to be very effective in the detection of LR attacks because it ensures that the largest loads in a system are monitored. Prior work on FDI attacks on SE shows that, to cause significant consequences, an attacker is required to target large loads in order to create large power flow changes \cite{Yuan2011,Zhang,Sanjab,Xie}.

\subsection{Detection algorithm}\label{detection_algorithm}
Dividing the $n$ system loads into groups allows us to overcome the dimensionality issue observed in Section \ref{largesystems}. The basic nearest neighbor detector can be used on large systems by running the algorithm individually on each load group. In this case, a threshold $\tau_j$ must be defined for each individual group $g_j$, for $j \in [1:n_g]$. 
The vector $\bm{p}\in \mathbb{R}^n$ containing the estimated loads computed by SE is divided into $n_g$ groups according to the procedure described in the previous subsection. Let us define $\bm{p}^j$ as the vector containing the real-time values of the loads in group $g_j$. For each group, the minimum distance between the load vector $\bm{p}^j$ and the corresponding loads in the historical dataset is calculated as
\begin{equation}
d_j=\min_{r=[1:n\textsubscript{\textit{h}}]} \|\bm{p}^j-\bm{h}^j_r\|\textsubscript{\textsubscript{2}}\label{distance}
\end{equation}
where $\bm{h}^j_r$ is the subset of loads belonging to group $g_j$ from the $r^{th}$ historical load vector. The minimum distance is then compared to the threshold $\tau_j$ to determine if the loads in group $g_j$ are normative or anomalous. Specifically, if $d_j >  \tau_j$, an alarm is raised, while if $d_j <  \tau_j$  the loads are considered attack-free. This process is repeated for every group and if one or more alarms are raised, the load vector $p$ is labelled as anomalous.

\section{Testing the improved detector}	\label{test}
\subsection{Experimental procedure}\label{procedure}
The performance of the nearest neighbor-based detector in conjunction with the grouping strategy is tested in depth in the following sections. The detection capability is measured both on intelligently designed attacks as well as random load redistribution attacks; moreover, we study its sensitivity to different parameters, such as the load shift of the attack and the number of attacked buses.

The goal of the following experiments is to analyze the quality of the detector at understanding if a load vector is normative or attacked.  
The primary test system used is the synthetic Texas system described in Section \ref{largesystems}; all numerical results discussed below are based on this system. Additional testing performed on the 2383 bus Polish test case \cite{matpower}, for which we generated historical load profiles based on real data from a major US ISO \cite{svd}, showed comparable results and it is here omitted due to space constraints.
	
First, the 1125 system loads in the Texas system are divided into groups following the procedure from Section \ref{grouping}. For the tests described below, the parameters chosen for the creation of the groups are: radius $r_g = 7$ and number of groups $n_g = 35$. These values ensure that more than 60\% of the loads in the system are included in one or more groups and the ones that are outside of the groups have at least one monitored neighboring load. Moreover, these load groups are equally spread across the system; as a result, the system is effectively monitored in its entirety. Preliminary testing has shown that increasing the number of groups (and, thus, of the loads considered) did not improve detection performance.

In each experiment, two datasets are needed: the normative load data $\bm{P}_N \in \mathbb{R}^{1125 \times 8784}$  and the anomalous load data $\bm{P}_A \in \mathbb{R}^{1125 \times H}$, where $H$ varies for the different type of attacks. The normative data represents one load vector for each hour of 2016 (2016 was a leap year). The set $\bm{P}_A$ contains attacked load vectors which are designed starting from the normative load vectors in $\bm{P}_N$; depending on the type of attack, some of the loads are modified either intelligently or randomly, as described below. 

To compute detection probability and false alarm, the load vectors of dataset $\bm{P}_N$ are first divided into three subsets: historical, training, and testing. The historical dataset $\bm{P}_N^\text{hist}$ includes 70\% of the total hours of 2016 and it represents the past loads known to the system operator and used in its nearest neighbor algorithms. The training dataset $\bm{P}_N^\text{train}$ represents another 20\% of $\bm{P}_N$ and it is needed to determine the thresholds $\tau_j$ for each load group. The remaining 10\% of normative load vectors is used as the testing dataset $\bm{P}_N^\text{test}$ to determine the false alarm rate. To determine the threshold $\tau_j$ for group $g_j$, the minimum distance $d_{i,j}$ between each load vector $\bm{p}^j_i$ for $i$ in $\bm{P}_N^\text{train}$ and the historical dataset is computed using \eqref{distance}. The threshold $\tau_j$ is defined as a fixed fraction of the maximum closest distance, defined for each group as
\begin{equation}
d_{\text{max},j}=\max_{p_i \in \bm{P}_N^\text{train}} d_{i,j}.
\end{equation}
For each load vector in $\bm{P}_N^\text{test}$, the minimum distance from $\bm{P}_N^\text{hist}$ is computed and compared with the threshold: the false alarm rate is the ratio between the number of times a load vector is labelled as attacked (e.g. at least one load group has minimum distance greater than its corresponding threshold) and the total number of load vectors in $\bm{P}_N^\text{test}$. Similarly, the minimum distance is calculated for every attacked case and the detection probability is computed. As we will explain in more detail in the next sections, varying the threshold about the value $d_{\text{max},j}$ allows to span different detection probabilities and false alarm rates in order to determine the receiver operating characteristic (ROC). The proposed algorithm is extremely efficient and testing a load vector only takes a fraction of a second on a normal laptop; thus, even on large power systems, the detector can easily run in real-time.

\begin{figure}
	\centerline{\includegraphics[scale=0.27]{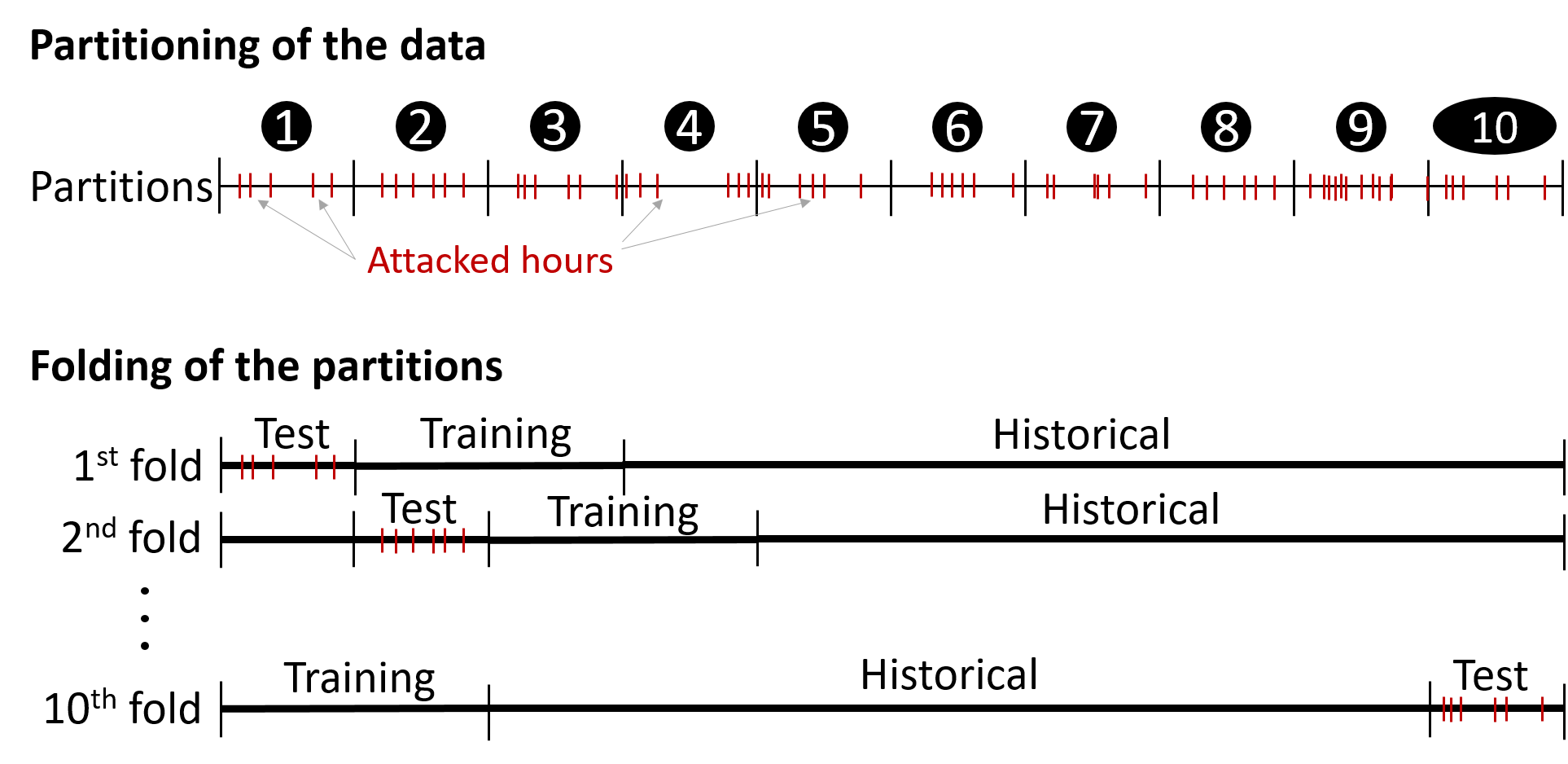}}
	\caption{Description of the 10-folding technique and definition of the datasets}
	\label{folding} 
\end{figure}

Because the normative load dataset is limited to one year, in order to have a more complete assessment of the performance of the detector, a $k$-folding technique is used to test every hour of the year by rotating through multiple sets of historical, training, and testing datasets. The hours of 2016 are randomly divided into ten equally sized partitions as illustrated in Fig.~\ref{folding}; the partitions are fixed throughout the testing process. For the first fold, the load vectors corresponding to the hours in the first partition are assigned to $\bm{P}_N^\text{test}$, the second and third partitions to $\bm{P}_N^\text{train}$ and the remaining seven represent the historical data $\bm{P}_N^\text{hist}$. Given these partitions, the number of false alarms and the number of attacks detected are calculated on the normative and attacked load vectors in the testing partition. The subsequent folds are created by shifting the partitions assigned to the three datasets by one: for example, in the second fold $\bm{P}_N^\text{test}$ will coincide with the second partition, $\bm{P}_N^\text{train}$ with the third and fourth partitions, and $\bm{P}_N^\text{hist}$ with the remaining ones. The final detection probability is then calculated by adding up the correctly identified attacks across all folds and dividing by the total number of attacks; the false alarm rate is the total number of false alarms divided by 8784.


\subsection{Detection of intelligently designed attacks}\label{intell}
We use the bi-level problem in \cite{Zhang} to design attacks that simulate specific changes in loads to cause physical overflows on a target line, while being unobservable to the system operators (and SE). The testing procedure described in the previous sections is employed here to verify the ability of the proposed detector and grouping strategy in correctly identifying malicious loads resulting from these intelligent attacks. 

The bi-level problem in \cite{Zhang} is structured so that any one branch can be selected as a target, and an attack will be designed to maximize the flow on it. Depending on the specific system conditions, a successful attack (that is, one causing the resulting power flow to go above the branch rating) may not exist; generally, the higher the pre-attack flow, the more likely the attack will lead to overflow. For this reason, the first step in designing the attacks is to run an AC optimal power flow (ACOPF) for every load vector in $\bm{P}_N$ to identify any congested branch. For the purpose of this study, a congested branch is any line or transformer that has a base case power flow loading of 90\% of its rating or more. Attacks are designed on each hour of 2016 for which one or more branches are congested. These branches are individually selected as targets of the attacks; thus, an hour will have as many different attacks as the number of branches with base case flow above 90\% in that hour. Moreover, for each target branch, attacks are designed with a load shift factor ranging from 1\% to 15\% in steps of 1\%. This allows us to study how the detection performance varies in relation to the attack magnitude. As a result of this process, 8861 successful attacks are computed, across every hour, target line, and load shifts. 

\begin{figure}
	\centerline{\includegraphics[scale=0.3]{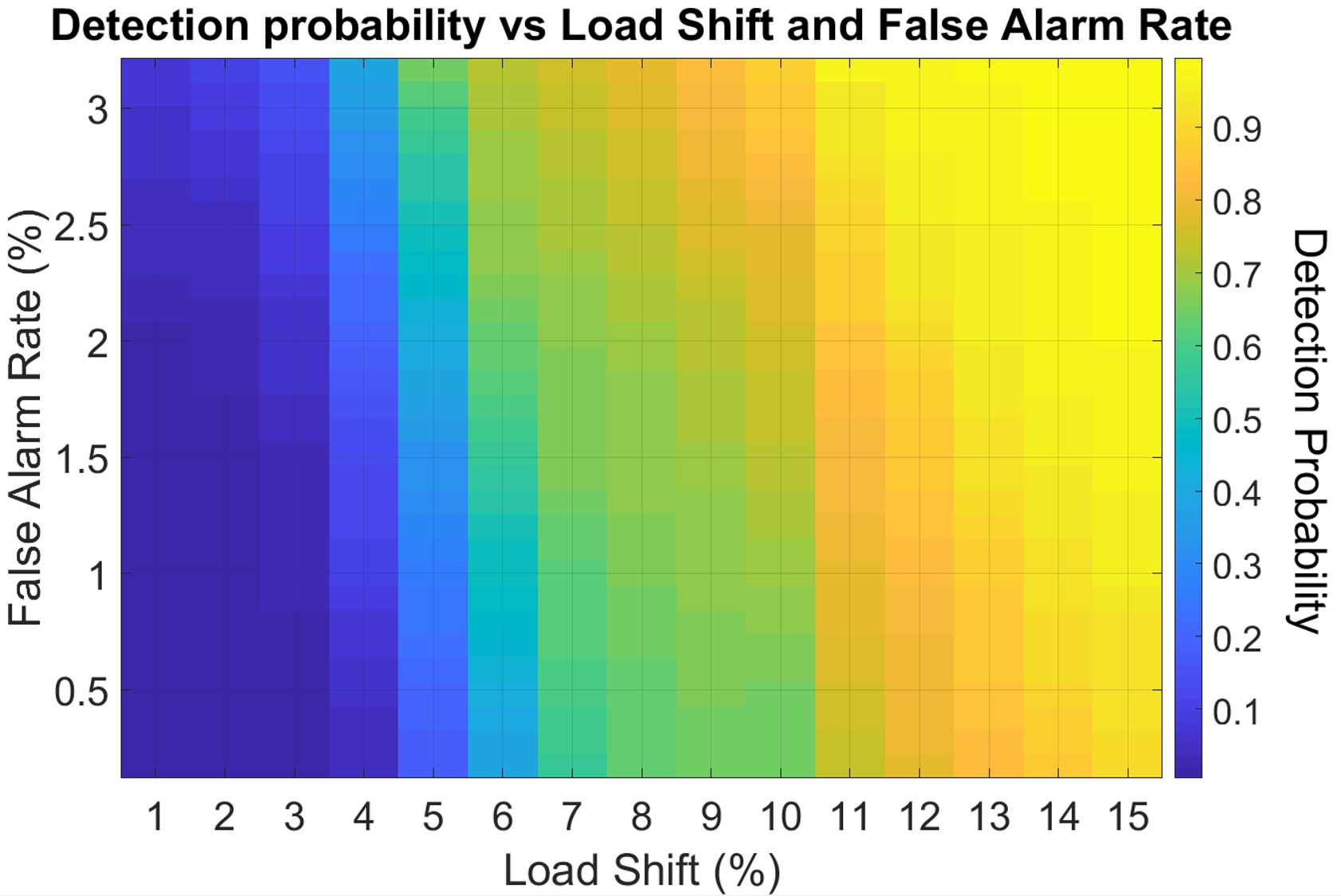}}
	\caption{Intelligent attacks: detection probability as a function of load shift and false alarm rate.}
	\label{real_DPvsLSandFA} 
\end{figure}

\begin{figure}
	\centerline{\includegraphics[scale=0.3]{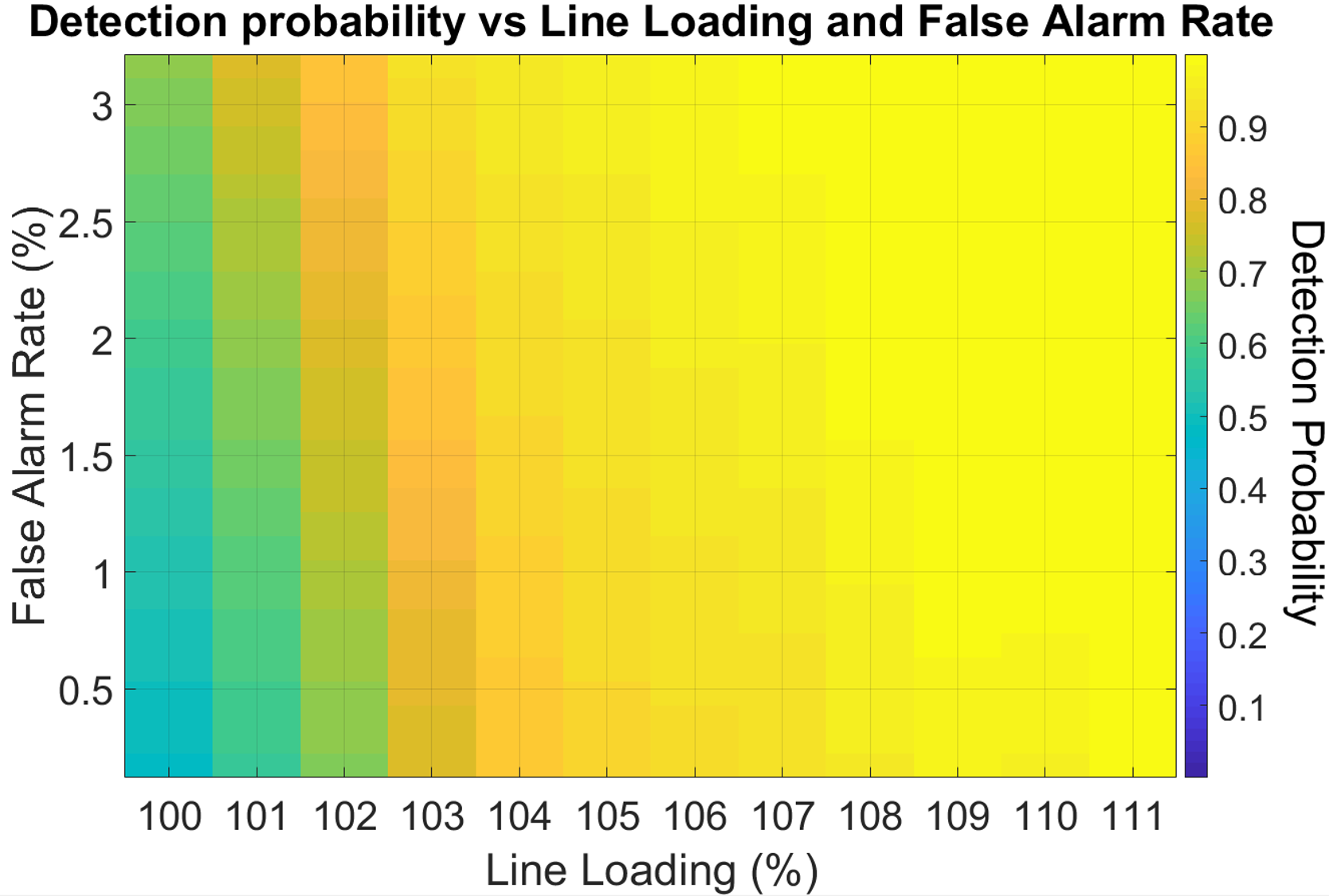}}
	\caption{Intelligent attacks: detection probability as a function of line overload and false alarm rate.}
	\label{real_DPvsOLandFA} 
\end{figure}

The resulting attacked load vectors have been tested following the $k$-folding procedure in Section \ref{procedure}, where the threshold for each group $g_j$ was varied from 0.9 to 1.1 times $d_{\text{max},j}$. Figure~\ref{real_DPvsLSandFA} shows the detection probability (colored scale) as a function of the load shift (x-axis) and the false alarm rate (y-axis). It can be seen that the detector does not perform well on attacks with very low load shifts, while for load shift between 10 and 15\% the detection probability goes from 80 to 100\% with false alarm rates ranging from 0 to 3\%. While the load shift factor is an important metric in the design phase of the attacks, from an operator's perspective it is more meaningful to evaluate the physical consequences of the attacks. Figure~\ref{real_DPvsOLandFA} shows the detection probability as a function of the line overload resulting from the attacks. From this figure we can easily see that the detector has extremely high probability of detecting any attack that would cause important physical damage: considering the safety margins built into the operational tools, an overload of 2 or 3\% is not likely to cause any system disruption.

\subsection{Detection of random load redistribution attacks}\label{random}
The experiments in the previous section have shown that the proposed detector is very effective in identifying attacked load vectors designed to create significant overflows on specific target lines. In this section we study the sensitivity of this algorithm to anomalous loads which have not necessarily been intelligently designed. To do so, a large number of false load vectors will be created starting from the historical data; the detection performance is then computed as the number of modified loads and the amount of load change are varied across a broad spectrum. 

\begin{figure}
	\centerline{\includegraphics[scale=0.29]{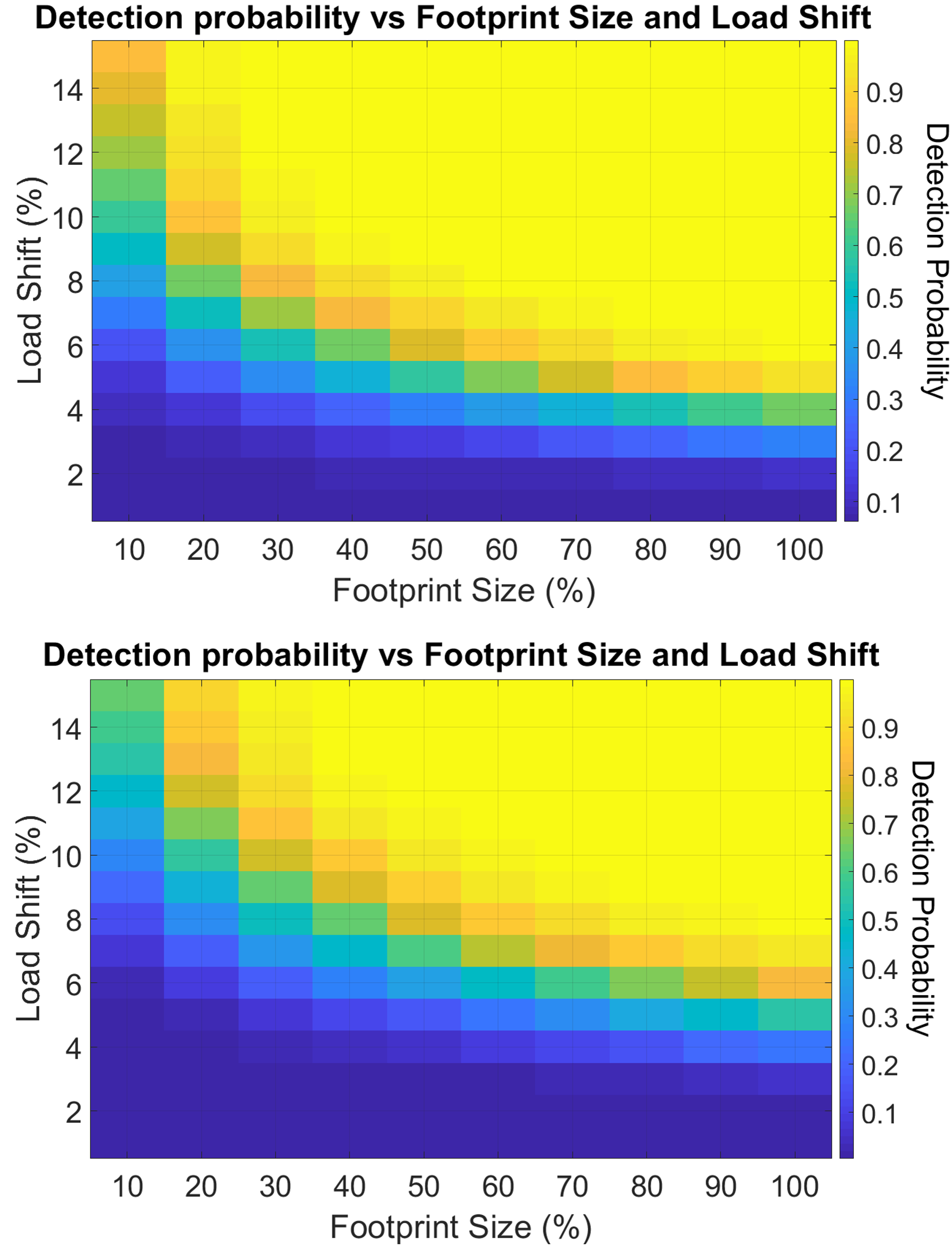}}
	\caption{Random attacks: detection probability as a function of load shift and footprint size for false alarm rate of 5.5\% (top) and 0.4\% (bottom).}
	\label{random_DPvsLSandPCT_high} 
\end{figure}

\begin{figure}
	\centerline{\includegraphics[scale=0.29]{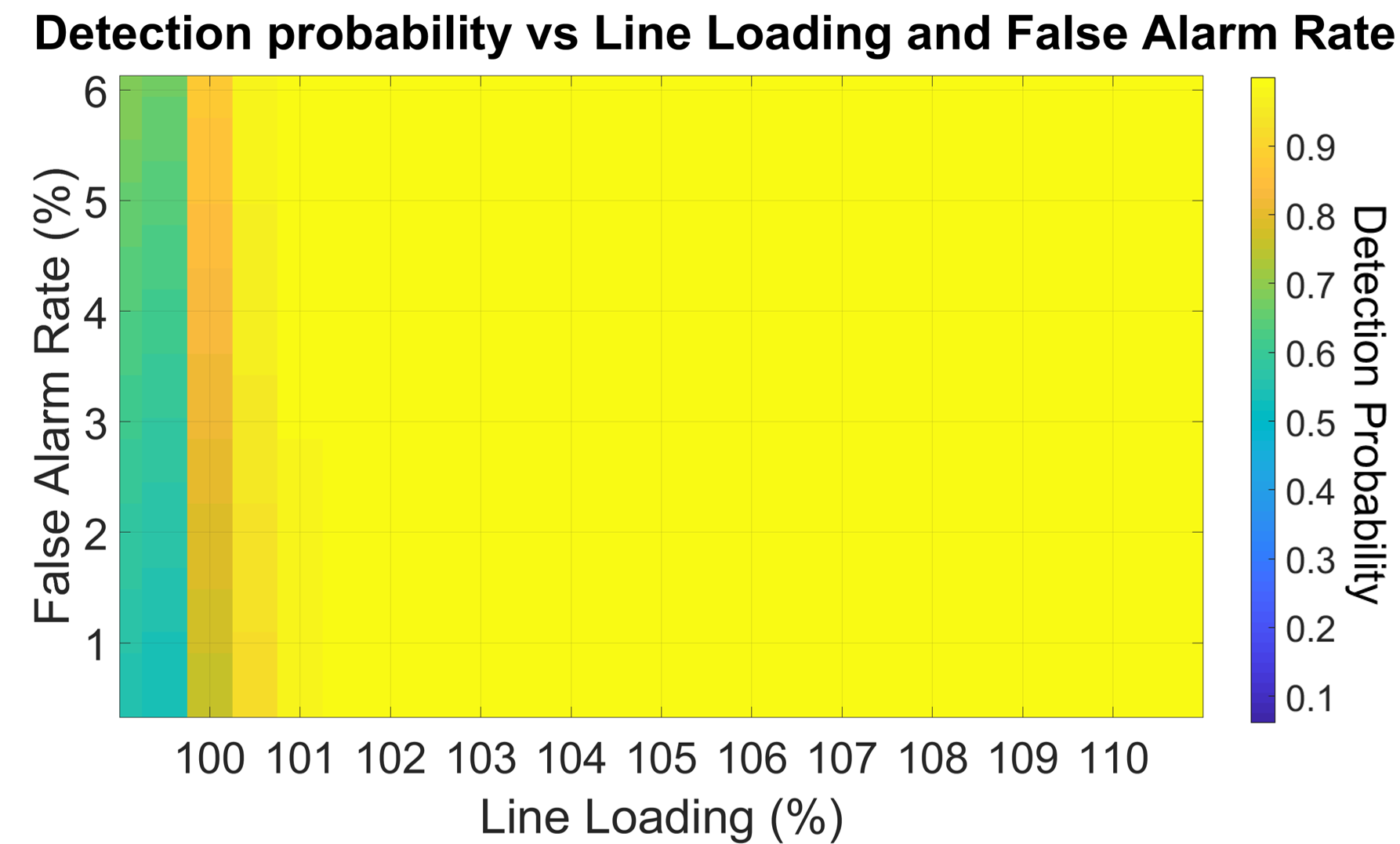}}
	\caption{Random attacks: detection probability as a function of line overload and false alarm rate.}
	\label{random_DPvsOLandFA} 
\end{figure}

The false load vectors are created by randomly selecting a subset of the loads in each vector of $\bm{P}_N$ and modifying them by either increasing or decreasing their value by a given load shift factor. For this study, the same load shifts as in the previous section are used, while the footprint size of the attack as a percentage of the total number of system loads is varied between 10\% and 100\% in steps of 10\% for every hour. The resulting anomalous load dataset $\bm{P}_A$ has dimensions $1125 \times H$, where $H = 8784 \times 15 \times 10 = 1,317,600$. Similarly to what done in the previous section, all these false load vectors are fed to the proposed detector and the detection probability computed. In this case, the detection probability is a function of three parameters: the false alarm rate, the load shift, and the footprint size. Figure~\ref{random_DPvsLSandPCT_high} shows the detection probability (colored scale) as a function of the footprint size (x-axis) and the load shift (y-axis), when a high false alarm rate of 5.5\% is allowed (top) and for a very low false alarm rate of 0.4\% (bottom). Clearly, for a given load shift and footprint size, the detection probability is higher when the false alarm rate is higher. Overall, the detector performs well, having perfect detection capability for a wide range of different attacks. Compared to the detection performance on intelligently designed attacks, the detector is not as good at identifying the random attacks with small load shift and small percentages of attacked loads. This can be explained by the fact that the intelligent attacks are designed in such a way that the modified loads belong to a spatially concentrated subgraph, thus it is likely that some of the load groups will include a large number of attacked loads. In the random attacks, the loads are modified across the whole network and, hence, distributed across a higher number of groups; because of this, each group will experience a smaller deviation from the normative data, resulting in worse detection capability. On the other hand, because of this fact, the random attacks are less likely to cause line overloads. Figure~\ref{random_DPvsOLandFA} shows the detection probability versus line overload and false alarm rate. 
From these results it can be seen that any random attack that would result in line overloads is easily detected, demonstrating the high effectiveness of the proposed algorithm in detecting anomalous and dangerous load vectors. 
	
%

\subsection{Integration within EMS}

\begin{figure*}
	\centerline{\includegraphics[scale=0.53]{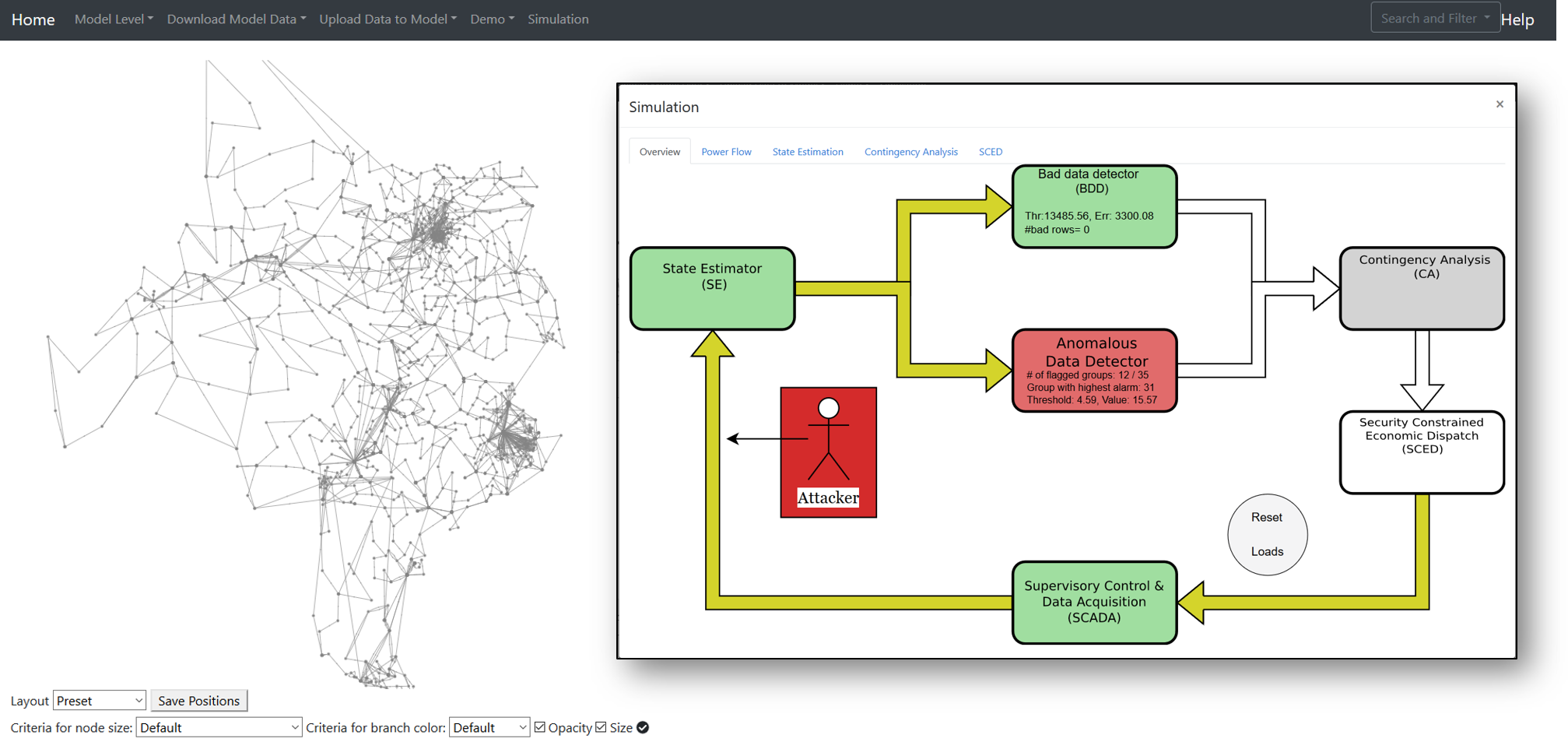}}
	\caption{Implementation of the improved bad data detector within an EMS}
	\label{EMS} 
\end{figure*}

The proposed detector has been fully implemented in a state-of-the-art EMS platform developed at Arizona State University \cite{SankarWebsite}. This software was created as part of NSF Grant 1449080, by an ASU team lead by Dr. Lalitha Sankar, Dr. Kory Hedman, and Dr. Oliver Kosut and in collaboration with Dr. Robin Podmore from IncSys, Inc., leader in power system simulation tools and operator training \cite{robin}, \cite{incsys}. Figure~\ref{EMS} shows the interface of the platform: on the left is the network graph of the Texas system, while, on the right, the simulation page with the main blocks of the EMS is displayed. In the example shown, the traditional residue-based BDD has easily been bypassed, while the proposed anomalous load detector identifies the attack and gives information on the extent of the attack based on the number of groups that raised a flag. Overall, this platform allows for the testing of the detector in a realistic power system operations environment while showcasing its effectiveness in terms of computational efficiency and integration within energy managements systems. Details on the design of the software platform and its building blocks can be found in \cite{Roozbeh}, while the code for the attack detection algorithm is freely available on Github \cite{git}.

\section{Attack localization}\label{localization}
In the previous sections, we have introduced a load anomaly detector based on nearest neighbor and a grouping strategy which was shown to have excellent performance against both intelligently designed attacks and random load changes. This algorithm can be extended beyond simply determining whether a load vector contains anomalous data or not; it can be leveraged to determine which buses have been modified or are deviating from their usual behavior. Localizing the subgraph affected by an attack or load anomaly represents a step forward in terms of system operations security. Knowing which loads are likely to have caused the detector to raise an alarm is an important step in the implementation of secure EMS functionalities. 
For example, the load values which are determined to be unreliable could be replaced by forecasted values or an uncertainty margin assigned to them so that the system could be operated in a secure state. 

A similar approach for secure operations against cyber-attacks is studied in \cite{Abusorrah2017}, where the authors present an optimal dispatch problem to find a secure and cost-effective dispatch solution considering variable bus loads and, thus, protecting the system from unexpected load changes. Also, in \cite{Shayan}, a secure unit commitment (UC) problem is formulated such that in case of a cyber attack the system operator can switch from the normal UC solution to a secure one while following all network constraints. The issue with these approaches is that it would cause the system to be operated in a too conservative and, thus, less efficient state for most of the time. The advantage of being able to detect and localize an attack is that the system operator can make a better informed decision on when and how to secure the system, without impacting normal operations.

\subsection{Likelihood determination}
The grouping strategy provides an approximate way of localizing the attacks by identifying groups of loads that deviate from their normative behavior. In this section, we describe a statistical approach to further analyze the values of the individual loads to identify which ones are more likely to have triggered the detector. Because of the many attack subgraphs that are possible, determining exactly which are the attacked loads would be extremely hard. For this reason, our goal is to assign to each load a probability $q_l$ that represents the likelihood of that load being attacked. In this sense, the likelihood is a risk measure and it can be quantified using an empirical metric that relies on estimated likelihoods, namely average log-loss (also known as cross-entropy) \cite{Cover}. Average log-loss is defined as
\begin{equation}\label{logloss}
l = \frac{1}{n_L} \sum\limits_{l=1}^{n_L} -[y_l \text{log}_2(q_l)+(1-y_l)\text{log}_2(1-q_l)]
\end{equation}
where $n_L$ represents the total number of samples (e.g. loads tested), $q_l$ is the probability associated with each load, and $y_l$ is 1 if the load was indeed attacked and 0 if it was not modified.

We define the values of the loads in group $j$ at time $i$ as $p_i^j = [p_{i,1}^j, p_{i,2}^j,\ldots, p_{i,k_j}^j]^T$,
where $k_j$ is the number of loads in group $j$. The minimum distance $d_{i,j}$ between the load vector $p_i^j$ and the historical data is computed using \eqref{distance}; as explained in Section \ref{detection_algorithm}, if $d_{i,j}$ is greater than threshold $\tau _j$, group $j$ is said to raise a violation at time $i$. Moreover, define the loads in the nearest neighbor of $p_i^j$ as $h_r^j = [h_{r,1}^j, h_{r,2}^j,\ldots, h_{r,k_j}^j]^T$,
where $h_r$ is the $r^{th}$ historical load vector in $\bm{P}_N^\text{hist}$.
For each load in group $j$, the normalized difference between load $l$ at time $i$ and its corresponding value in the nearest neighbor $h_r^j$ is computed as
\begin{equation}\label{delta}
\delta _{i,l}^j = \left| \frac{p_{i,l}^j - h_{r,l}^j}{h_{r,l}^j} \right| \qquad l=1,\ldots,k_j.
\end{equation}
We cannot directly look at this normalized difference to know if a load is attacked because different loads could have different amounts of deviation. In order to account for this variability, we determine the normative behavior of each load by computing the first and second order statistics of its normalized difference:
\begin{equation}
\mu _{	\delta _l^j} = \frac{1}{n_i} \sum\limits_{i \in \bm{P}_N^\text{train}} \delta _{i,l}^j \qquad \forall l, \forall j
\end{equation}
\begin{equation}
\sigma _{	\delta _l^j} = \sqrt{\frac{1}{n_i} \sum\limits_{i \in \bm{P}_N^\text{train}} (\delta _{i,l}^j - \mu _{	\delta _l^j})^2} \qquad \forall l, \forall j.
\end{equation}
Given a specific load vector $p_i \in {P}_N^\text{test}$ and its corresponding $\delta _{i,l}^j$ for all $l$ and $j$, we determine how far each load deviates from the normative behavior using a Z-score which is defined as follows:
\begin{equation}\label{zscore}
z_{i,l}^j = \frac{\delta _{i,l}^j - \mu_{\delta_l^j}}{\sigma _{	\delta _l^j}} .
\end{equation}
Intuitively, the Z-score indicates the number of standard deviations by which $\delta _{i,l}^j$ is above (or below) the mean for load $l$ in group $j$ observed in the attack-free data.	

Based on this setup, there exists a joint distribution $Q(a,v,z)$ between whether a load was attacked ($a=1$) or not ($a=0$), if it belongs to a group that raised a violation ($v=1$) or not ($v=0$), and its Z-score $z$. While $Q(a,v,z)$ is not known, we can empirically estimate the conditional probability $Q(a|v,z)$ of a load being attacked given its Z-score and whether it raised a violation or not. In other words, our goal is to define a likelihood function $\mathcal{L}(v,z)$ that takes as inputs a load's Z-score and whether it raised a violation to determine the probability that the load is attacked. 
\begin{figure}
	\centerline{\includegraphics[scale=0.28]{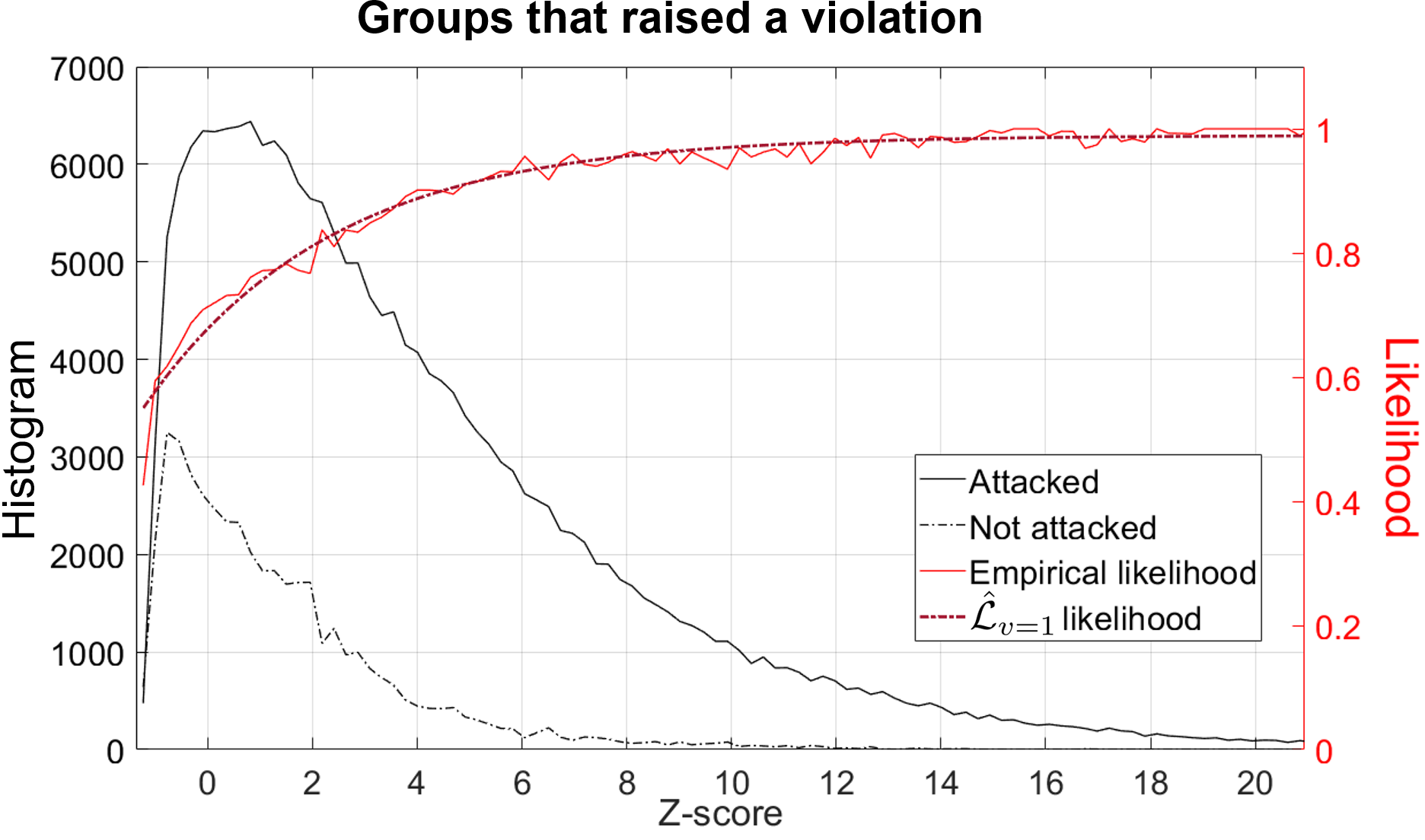}}
	\caption{Distribution of Z-scores (in black) and likelihood function (in red) for loads in groups that raise a violation.}
	\label{loc_like_vio} 
\end{figure}
\begin{figure}
	\centerline{\includegraphics[scale=0.29]{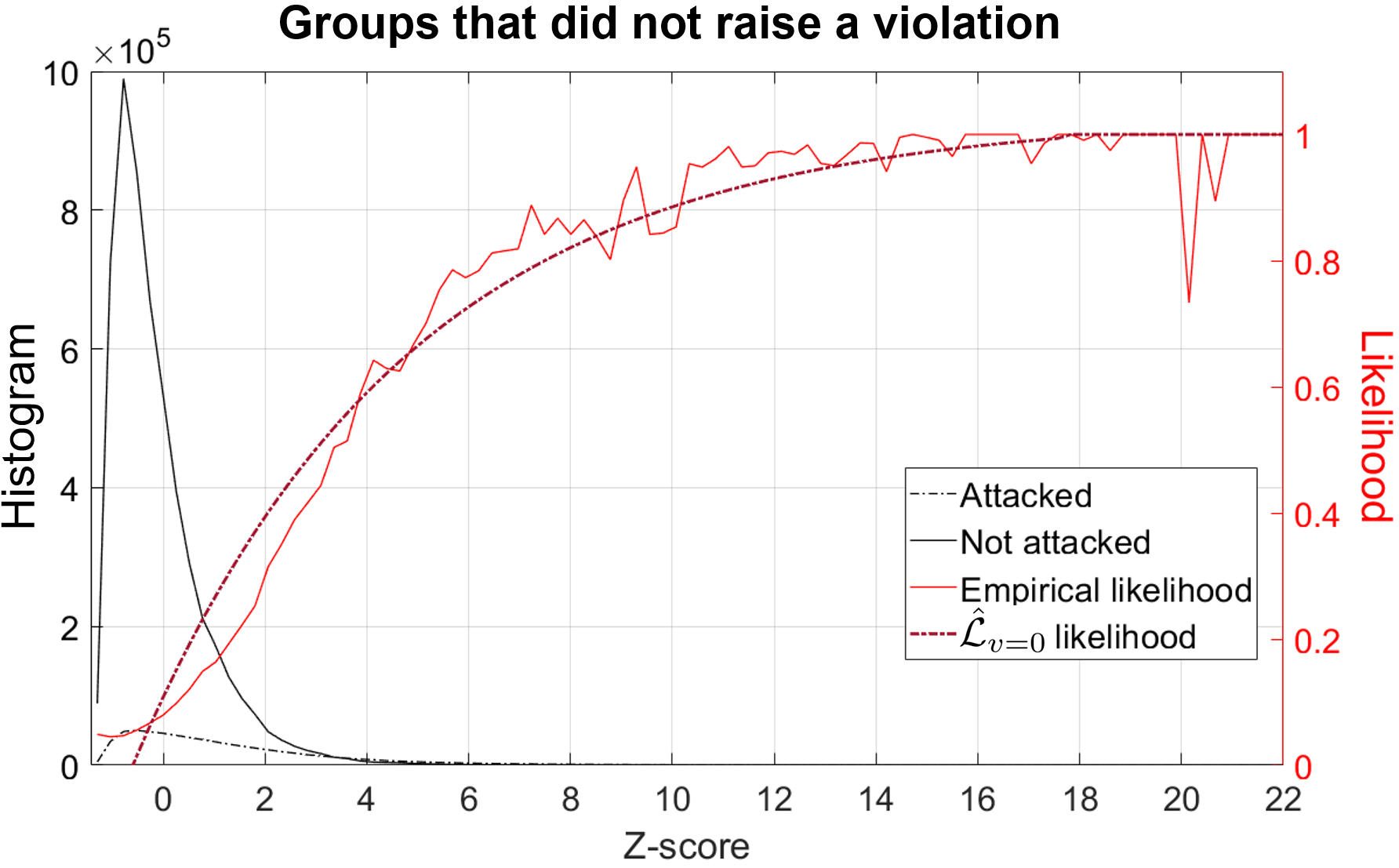}}
	\caption{Distribution of Z-scores (in black) and likelihood function (in red) for loads in groups that do not raise a violation.}
	\label{loc_like_notvio} 
\end{figure}

First, we compute the Z-score \eqref{zscore} for all intelligently designed attacks in ${P}_A$ that result in an overload of 3\% or more. As discussed in Section \ref{intell}, those are the attacks that can cause significant damage and they are almost always detected by the nearest neighbor algorithm. The histogram of the Z-score for the loads that belonged to groups that raised a violation is shown in Fig.~\ref{loc_like_vio}. In particular, the solid black line represents the histogram of Z-scores for loads that were attacked and we indicate as $\phi _{a=1,v=1}(z)$, while the black dotted line represents the histogram  for loads that were not attacked $\phi _{a=0,v=1}(z)$. From these two curves, we notice that overall if a load belongs to a group that raised a violation it is very likely that the load is indeed being attacked. Moreover, the higher the Z-score, the more likely it is that a load is attacked. Based on these observations we can now define a function that maps the Z-score of a load to the likelihood of the load being attacked. The estimated conditional likelihood for loads that are in groups that raise a violation is computed as 
\begin{equation}\label{empiricallikelihood}
\bar{\mathcal{L}}_{a|v=1,Z=z} = \frac{\phi _{a=1,v=1}(z)}{\phi _{a=1,v=1}(z)+\phi _{a=0,v=1}(z)}
\end{equation}
and it is shown by the solid red line in Fig.~\ref{loc_like_vio}. For the set of data points we obtain using \eqref{empiricallikelihood}, we fitted a smooth curve of the form $a\cdot e^{-b\cdot x}+c$ to avoid overfitting as shown by the dotted red line. This curve is defined as the conditional likelihood $\hat{\mathcal{L}}_{a|v=1,Z=z}$ which can be used to assign to each load a probability of being attacked based on its Z-score. The same procedure is performed on the loads in groups that do not raise a violation and the corresponding likelihood $\hat{\mathcal{L}}_{a|v=0,Z=z}$ is estimated; the results are shown in Fig.~\ref{loc_like_notvio}. Comparing the two conditional likelihood functions we notice that, for low Z-score values, $\hat{\mathcal{L}}_{a|v=1,Z=z}$ reaches a minimum likelihood value of around 0.5 while $\hat{\mathcal{L}}_{a|v=0,Z=z}$ reaches zero.

\subsection{Numerical results}
The performance of this approach is tested on the intelligently designed attacks from Section \ref{intell}, with $\tau_j = d_{\text{max},j}$. The conditional likelihood functions $\hat{\mathcal{L}}_{a|v=1,Z=z}$ and $\hat{\mathcal{L}}_{a|v=0,Z=z}$ are learned on 70\% of the attacks and they are tested on the remaining 30\%. The Z-score for every load is computed using \eqref{zscore} and the average log-loss as in \eqref{logloss}.

As a way of comparison, we also tested two simpler approaches to assign likelihood values to each load. The first one does not rely on the Z-score and only considers if the load belongs to groups with violations or not: based on our data, on average, in a group that raised a violation 82\% of the loads are attacked, while in groups that did not raise violations only 10\% are actually attacked. Based on this prior knowledge, the first simple approach assigns a likelihood $q_l = 0.82$ if load $l$ is in a group with violations and $q_l = 0.10$ otherwise. The second approach is even simpler and it assigns a fixed likelihood to every load regardless of which group they belong to; from our results, the optimal value for this approach is $q_l=0.15$. The results of the Z-score-based approach (indicated as $q(a|v,z)$), as well as the two simpler ones (indicated as $q(a|v)$ and $q(a)$) are summarized in Table \ref{table_loss}. We can see that the more sophisticated the approach (i.e. the more information is used), the smaller the average log-loss.

\renewcommand{\arraystretch}{1.5}
\begin{table}[h]
	\centering
	\caption{Performance comparison of the three likelihood approaches.}
	\begin{tabular}{| c | c | c | c |}
		\hline
		\textbf{Approach} & $q(a|v,z)$ & $q(a|v)$ & $q(a)$ \\ \hline
		\textbf{Average log-loss} & 0.340 & 0.489 & 0.608 \\ \hline	
	\end{tabular}		
	\vspace{-4mm}
	\label{table_loss}
\end{table}
\renewcommand{\arraystretch}{1}	

\section{Conclusion}
In this paper we presented an improved data-driven algorithm for the detection of load redistribution attacks and a statistical approach for the localization of the attacked buses. The detector, based on nearest neighbor and a grouping strategy, was tested on a large number of attacks belonging to two different classes: intelligent attacks and random load changing attacks. The results obtained on the synthetic Texas system show excellent detection capability, especially against the attacks that have the worst consequences on the system. The attack localization scheme assigns a likelihood value to each load indicating the probability of that load being attacked; this approach offers operators a greater insight in case of cyber-attacks allowing for more secure system operation.

As part of our future work, we intend to extend the detection algorithm to the analysis of different anomalies; the model can be trained to not only detect an anomaly, but also determine the type of event that caused it (cyber-attack, natural event, fault, etc.). Moreover, the algorithm can be enhanced by taking into consideration additional information about rare and sporadic events, such as forecasts of extreme weather events or temporary changes in load patterns due to known causes (e.g. sporting events, holidays, etc.). This could result in both improved detection probability and lower false alarm rate.

\bibliography{AttDet}

\begin{thebibliography}{10}
\providecommand{\url}[1]{#1}
\csname url@samestyle\endcsname
\providecommand{\newblock}{\relax}
\providecommand{\bibinfo}[2]{#2}
\providecommand{\BIBentrySTDinterwordspacing}{\spaceskip=0pt\relax}
\providecommand{\BIBentryALTinterwordstretchfactor}{4}
\providecommand{\BIBentryALTinterwordspacing}{\spaceskip=\fontdimen2\font plus
\BIBentryALTinterwordstretchfactor\fontdimen3\font minus
  \fontdimen4\font\relax}
\providecommand{\BIBforeignlanguage}[2]{{%
\expandafter\ifx\csname l@#1\endcsname\relax
\typeout{** WARNING: IEEEtran.bst: No hyphenation pattern has been}%
\typeout{** loaded for the language `#1'. Using the pattern for}%
\typeout{** the default language instead.}%
\else
\language=\csname l@#1\endcsname
\fi
#2}}
\providecommand{\BIBdecl}{\relax}
\BIBdecl

\bibitem{Liu2009}
Y.~Liu, P.~Ning, and M.~K. Reiter, ``{False data injection attacks against
  state estimation in electric power grids},'' \emph{Ccs}, vol.~14, no.~1, pp.
  1--33, 2009.

\bibitem{Yuan2011}
Y.~Yuan, Z.~Li, and K.~Ren, ``{Modeling Load Redistribution Attacks in Power
  Systems},'' \emph{IEEE Transactions on Smart Grid}, 2011.

\bibitem{Zhang}
J.~Zhang and L.~Sankar, ``Physical system consequences of unobservable
  state-and-topology cyber-physical attacks,'' \emph{IEEE Transactions on Smart
  Grid}, vol.~7, no.~4, pp. 2016--2025, 2016.

\bibitem{Sanjab}
A.~Sanjab and W.~Saad, ``{Data Injection Attacks on Smart Grids With Multiple
  Adversaries: A Game-Theoretic Perspective},'' \emph{IEEE Transactions on
  Smart Grid}, vol.~7, no.~4, pp. 2038--2049, jul 2016.

\bibitem{Xie}
L.~Xie, Y.~Mo, and B.~Sinopoli, ``{Integrity Data Attacks in Power Market
  Operations},'' \emph{IEEE Transactions on Smart Grid}, vol.~2, no.~4, pp.
  659--666, dec 2011.

\bibitem{Pinceti}
A.~Pinceti, L.~Sankar, and O.~Kosut, ``{Load Redistribution Attack Detection
  using Machine Learning: A Data-Driven Approach},'' in \emph{IEEE Power and
  Energy Society General Meeting}, 2018.

\bibitem{Joshi2017}
V.~Joshi, J.~Solanki, and S.~K. Solanki, ``{Statistical methods for detection
  and mitigation of the effect of different types of cyber-attacks and
  parameter inconsistencies in a real world distribution system},'' in
  \emph{2017 North American Power Symposium, NAPS 2017}, 2017.

\bibitem{Yu2018}
J.~J. Yu, Y.~Hou, and V.~O. Li, ``{Online False Data Injection Attack Detection
  with Wavelet Transform and Deep Neural Networks},'' \emph{IEEE Transactions
  on Industrial Informatics}, 2018.

\bibitem{Huang2016}
Y.~Huang, J.~Tang, Y.~Cheng, H.~Li, K.~A. Campbell, and Z.~Han, ``{Real-time
  detection of false data injection in smart grid networks: An adaptive CUSUM
  method and analysis},'' \emph{IEEE Systems Journal}, 2016.

\bibitem{Chu}
Z.~Chu, J.~Zhang, O.~Kosut, and L.~Sankar, ``Evaluating power system
  vulnerability to false data injection attacks via scalable optimization,''
  \emph{Smart Grid Communications (SmartGridComm), 2016 IEEE International
  Conference}, Nov. 2016.

\bibitem{nn1}
T.~Cover and P.~Hart, ``Nearest neighbor pattern classification,'' \emph{IEEE
  Transactions on Information Theory}, vol.~13, no.~1, January 1967.

\bibitem{nn2}
V.~Chandola, A.~Banerjee, and V.~Kumar, ``Anomaly detection : A survey,''
  \emph{ACM Computing Surveys}, vol. 41(3), no.~15, July 2009.

\bibitem{PJM}
\BIBentryALTinterwordspacing
PJM, ``{P}{J}{M} metered load data.'' [Online]. Available:
  \url{https://dataminer2.pjm.com/feed/hrlloadmetered/definition.}
\BIBentrySTDinterwordspacing

\bibitem{Birchfield}
A.~B. Birchfield, T.~Xu, K.~M. Gegner, K.~S. Shetye, and T.~J. Overbye, ``{Grid
  Structural Characteristics as Validation Criteria for Synthetic Networks},''
  \emph{IEEE Transactions on Power Systems}, vol.~32, no.~4, pp. 3258--3265,
  Jul 2017.

\bibitem{Li2018}
H.~Li, A.~L. Bornsheuer, T.~Xu, A.~B. Birchfield, and T.~J. Overbye, ``{Load
  modeling in synthetic electric grids},'' in \emph{2018 IEEE Texas Power and
  Energy Conference, TPEC 2018}, 2018.

\bibitem{matpower}
R.~D. {Zimmerman}, C.~E. {Murillo-Sánchez}, and R.~J. {Thomas}, ``Matpower:
  Steady-state operations, planning, and analysis tools for power systems
  research and education,'' \emph{IEEE Transactions on Power Systems}, vol.~26,
  no.~1, pp. 12--19, Feb 2011.

\bibitem{svd}
A.~Pinceti, L.~Sankar, and O.~Kosut, ``{Data-Driven Generation of Synthetic
  Load Datasets Preserving Spatio-Temporal Features},'' in \emph{IEEE Power and
  Energy Society General Meeting}, 2019.

\bibitem{SankarWebsite}
\BIBentryALTinterwordspacing
L.~Sankar, O.~Kosut, and K.~Hedman, ``A verifiable framework for cyber-physical
  attacks and countermeasures in a resilient electric power grid.'' [Online].
  Available: \url{https://sankar.engineering.asu.edu/nfs-dhs-cps-framework/}
\BIBentrySTDinterwordspacing

\bibitem{robin}
R.~Podmore, ``Digital computer analysis of power system networks,'' \emph{PhD
  Thesis, University of Canterbury, Christchurch, New Zealand}, 1972.

\bibitem{incsys}
\BIBentryALTinterwordspacing
{IncSys, Inc.}, ``Inc{S}ys - power system simulation software.'' [Online].
  Available: \url{www.incsys.com}
\BIBentrySTDinterwordspacing

\bibitem{Roozbeh}
\BIBentryALTinterwordspacing
R.~Khodadadeh, ``\BIBforeignlanguage{eng}{Designing a software platform for
  evaluating cyber-attacks on the electric power grid},''
  \emph{\BIBforeignlanguage{eng}{Master's Thesis}}, 2019. [Online]. Available:
  \url{http://search.proquest.com/docview/2228212793/}
\BIBentrySTDinterwordspacing

\bibitem{git}
A.~Pinceti, ``Nearest neighbor attack detection,''
  \url{https://github.com/apince/EMS_FDI_NearestNeighborAttackDetection}, 2019.

\bibitem{Abusorrah2017}
A.~Abusorrah, A.~Alabdulwahab, Z.~Li, and M.~Shahidehpour, ``{Minimax-Regret
  Robust Defensive Strategy Against False Data Injection Attacks},'' 2017.

\bibitem{Shayan}
H.~Shayan and T.~Amraee, ``{Network Constrained Unit Commitment Under Cyber
  Attacks Driven Overloads},'' \emph{IEEE Transactions on Smart Grid}, p.~1,
  2019.

\bibitem{Cover}
T.~M. Cover and J.~A. Thomas, \emph{Elements of information theory}.\hskip 1em
  plus 0.5em minus 0.4em\relax Wiley-Interscience, 1991, ch.~13.

\end{thebibliography}
\bibliographystyle{ieeetran}

\end{document}